\documentclass[preprint,3p,twocolumn]{elsarticle}
\usepackage{graphicx,t1enc}
\usepackage{amssymb,amsmath}
\usepackage{cuted}
\usepackage[colorlinks=true,allcolors=blue]{hyperref}

\journal{Chaos, Solitons \& Fractals}

\begin{document}

\begin{frontmatter}
\title{Allee effect in models of interacting species }

\author[cab]{Marcelo N. Kuperman}
\ead{kuperman@cab.cnea.gov.ar}

\author[cab]{Guillermo Abramson}
\ead{abramson@cab.cnea.gov.ar}

\affiliation[cab]{
            organization={Consejo Nacional de Investigaciones Cient\'{\i}ficas y T\'ecnicas, Centro At\'omico Bariloche (CNEA) and Instituto Balseiro},
            city={R8402AGP Bariloche},
            country={Argentina}}

\begin{abstract}
The search for more realistic models for interacting species has produced many adaptations of the original Lotka-Volterra equations, such as the inclusion of the Allee effect and the different Holling's types of functional response. In the present work we show that a correct implementation of both ideas together requires a careful formulation. We focus our work in the fact that a density dependent carrying capacity, combined with the Allee effect, can lead to meaningless effects. We illustrate the difficulties in predator-prey and two-species competition models, together with our proposed solution of the careful inclusion of the corresponding cubic terms.
\end{abstract}


\end{frontmatter}

\section{Introduction}

In his works published in the 1930s \cite{allee31}, W. C. Allee suggested the possibility that individuals in a population might benefit from the presence of conspecifics, implying that, in some cases, instead of intra-competition there could be a positive feedback. The phenomenon, that was later called \textit{Allee effect}, does not have a  clear definition and has been ambiguously used in many examples \cite{steph99}. 
In most cases, the Allee effect accounts for a positive correlation between population density and individual fitness. 
One of the most usual interpretations is that individuals may experience a difficulty mating when the population density drops below a certain level.

There are many scenarios where the   Allee effect can be observed.  The situations in which the benefits of conspecific presence may be relevant include dilution or saturation of predators, surveillance and defense, cooperative predation, social thermoregulation, etc. While the general rule is that the Allee effect occurs in small or scattered populations, examples of its occurrence at high population densities have been reported for some species \cite{courchamp08}.

The positive relationship between fitness and population size may be associated with a variety of mechanisms that affect reproduction and survival. As mentioned above, a well-established but not exclusive example is the limitation to find a mate, which can reduce the birth rate and lead to a population collapse in species with sexual reproduction. If reproduction, feeding and protection are cooperative to some extent, they will become more efficient in larger groups, with ensuing greater reproductive success or survival \cite{courchamp99}

There exist still other situations where there are not cooperative behaviors, yet  the presence of conspecific individuals is beneficial. For example, the risk of per capita predation is lower in larger prey populations than in small ones \cite{dennis89,gascoigne04}.  This effect can be included as a saturation in the predation term. In the present work, we consider this possibility. 

The ubiquity of different situations liable to be related to the Allee effect, as well as the lack of an accurate definition, has led to the considerations of two versions of the effect, defined by Stephens et al.~\cite{steph99} as the \textit{component Allee effect} and the \textit{demographic Allee effect}. The difference between them is that, in the former, the increase of the population affects some particular components of the fitness of an individual; in the latter, the effect manifests at the level of total fitness. These definitions clearly imply that the demographic Allee effect requires the existence of at least one component Allee effect, while the reciprocal is not true.

Summarizing the previous ideas,   it is possible to say that the Allee effect is a positive association between absolute average individual aptitude and population size. Such a positive association may result in a critical population size below which the population cannot persist \cite{steph99}.  When the Allee effect is responsible for the existence of such a threshold it is called \textit{strong}, otherwise, it is called \textit{weak}.

One way to capture both effects in a dynamical model of a population is by means of the following simple equation, which unlike the logistic has two stable equilibria. One of them is analogous to the stable equilibrium of the logistic equation, associated with saturation, while the other one corresponds to the possibility of extinction of below-threshold populations:
\begin{equation}
\frac{dx}{dt}= r\,x\left(x-a\right)\left(1-\frac{x}{K}\right).
\label{eq:allee}
\end{equation}
This equation is analogous to the voltage equation proposed by Nagumo \cite{nagumo1962} for the active transmission of a pulse along a nerve axon, and we will call it ``Nagumo model'' henceforth. It has three equilibria, $x=0$, $x =K$ and $x=a$. If we plot the expression on the right-hand side of Eq.~(1) we can see that, when $0<a<K$, the first two equilibria are stable,  while the third one is unstable. This corresponds to the strong Allee effect: a population smaller than $a$ becomes extinct. In these cases, $a$ is a critical value, a
threshold below which the population cannot persist. Extinction, unlike what happens with the logistic equation, is a stable situation. 
\begin{figure}
\includegraphics[width=8cm,height=6cm]{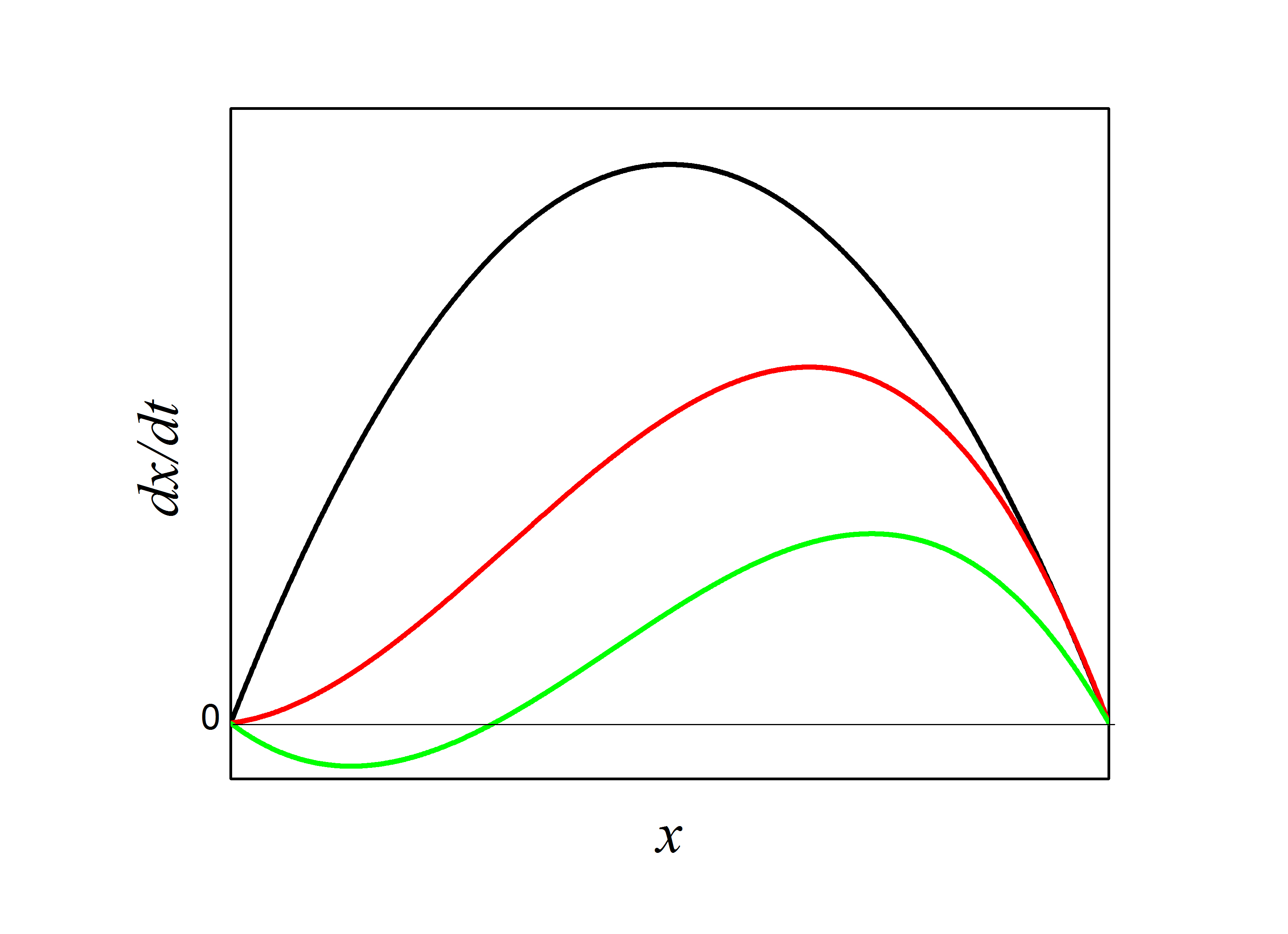}
\caption{Comparison between the logistic (black), weak Allee (red) and strong Allee (green).}
\label{fig:Allee}
\end{figure}
Negative values of $a$ add an equilibrium for negative $x$, which is irrelevant in the context of populations. However, this case serves to model the weak Allee effect, which is reflected in a slight departure from the logistic case, with a decrease in the growth rate for small values of $x$, but without stabilizing the extinction. Figure~\ref{fig:Allee} shows a comparison between the behavior of the right-hand side of Eq.~(\ref{eq:allee}) for both positive and negative values of $a$ and the logistic case. 

On the other extreme, if $a>K$ the resulting equation is meaningless.
There is no doubt that, for given values of the threshold and the carrying capacity, the Nagumo model can adequately describe the Allee effect in many situations. But, what if (as in~\cite{abramson2013}) the carrying capacity has its own dynamics? Such a case can arise in a predator-prey model where the carrying capacity of the predator is determined by the abundance of the prey. The Nagumo model for the predators would read: 
\begin{equation}
\dot{x}= r\,x(x-a)\left(1-\frac{x}{K(y)}\right),
\label{allee1}
\end{equation}
where the threshold $a$ is a characteristic feature of the species and is constant, but the carrying capacity $K(y)$ is defined by the population of the prey, $y$.  For the model to have the biological meaning that we expect, we need $a<K(y)$, implying that the carrying capacity must be larger than the critical value of the population to survive. Inasmuch as $y$ is a variable with its own dynamics, this relation cannot be assured. In such a case, the dynamics described by the Eq.~(\ref{allee1}) turns nonsensical: the population should get extinct (since the carrying capacity is smaller than the threshold), but at $K(y)=a$ the systems suffers a transcritical bifurcation, the equilibria interchange stability, and $x=a$ becomes the new stable equilibrium. In such a scenario, the whole equation loses its original biological meaning.

An analogous situation occurs when the competition of two species feeding on the same resource alters the carrying capacity of the environment relative to each species in a dynamical way dependent on the abundance of the competitor.
In such a case, a simple model for each species and in the presence of the Allee effect  is \cite{murray}:
\begin{equation}
\dot{x_i}= r_i x_i (t_i-x_i) \left(1-\frac{x_i}{K_i}-a_{ij}\frac{x_j}{K_i} \right),\\
\label{eq:comp1}
\end{equation}
where $i=1,2$ identify the competing species, and $t_i$ is the survival threshold for $x_i$. These equations  present no conflict provided that $(1-a_{ij} x_j/K_i)>t_i$.
But the same argument as in the case of the predator holds: what would happen if the carrying capacity associated with $x_i$ is reduced by the abundance of the competitor $x_j$, to values below its survival threshold? The model~(\ref{eq:comp1}) would no longer describe the real dynamics of the system, introducing new unrealistic stable equilibria. Furthermore,  the exchange of stability between the equilibria associated with the carrying capacity and survival thresholds can induce artificial oscillations of both competing populations, as will be shown below. 

In order to deal with all these issues rooted in an erroneous formulation of the equations describing the dynamics, we propose here a proper formalism that not only leads to correct results but also encompasses a sensible interpretation of ecological reality.

\section{Allee model for a dynamical carrying capacity}

Here we present an alternative mathematical formulation, one that preserves the spirit of the Allee effect in the context of a dynamically changing carrying capacity. Observe again Eq.~(\ref{allee1}). What we need, in order to solve the mentioned  problem, is that instead of a transcritical bifurcation when $K=a$, a saddle-node bifurcation occurs. If so, the stable equilibrium $x=K$ and the unstable one, $x=a$, collide and disappear. Meanwhile, the other equilibrium, $x=0$, must survive and remain stable.

We propose the following normal form, that satisfies the required conditions:
\begin{equation}
\dot{x}= \frac{x}{K} \left((K-a) |K-a|-(2x-a-K)^2\right)  
\label{allee2}
\end{equation}

The polynomial on the r.h.s. has three roots:
\begin{align*}
x_1 &= 0, \\
x_2 &= 1/2\left(a+K-\sqrt{K-a} \sqrt{\left| a-K\right|}\right), \\
x_3 &= 1/2\left(a+K+\sqrt{K-a} \sqrt{\left| a-K\right|}\right). 
\end{align*}

\begin{figure}[t]
\includegraphics[width=8cm,height=6cm]{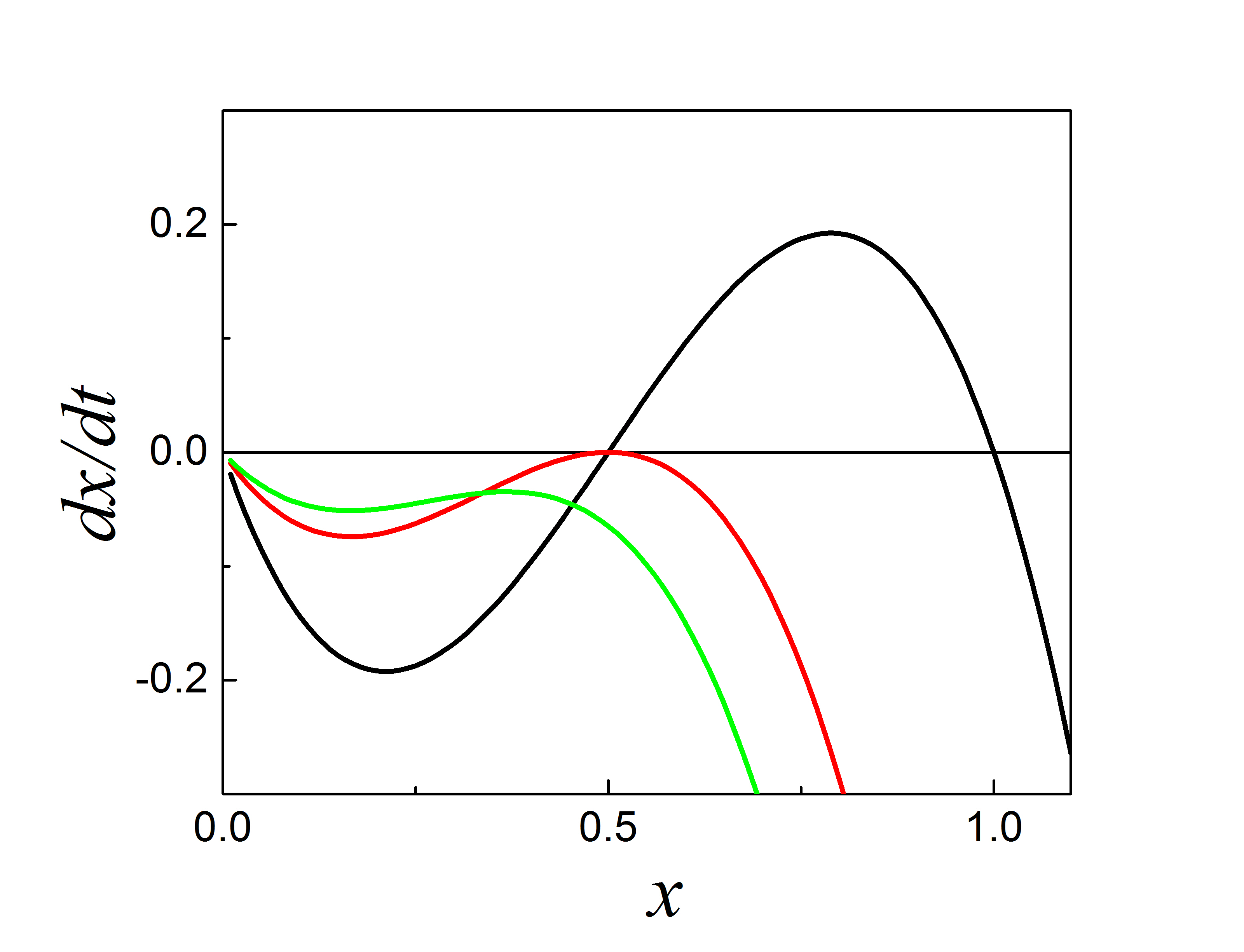}
\caption{Examples of the behavior of the r.h.s. of Eq.~(\ref{allee2}), for $K>a$ (black), $K=a$ (red), $K<a$ (green).}
\label{polroot}
\end{figure}

When $a<K$ the three roots are real: $x_1=0$, $x_2=a$, $x_3=K$, while when $a>K$ only $x_1=0$ is real. Figure~\ref{polroot} shows three examples of the polynomial behavior for $K\gtreqqless a$.

So far the formulation of a consistent scheme to include the consequences of the Allee term could be considered as a mere exercise. In order to assess its relevance to the dynamics, we compare the behavior of the present system with the one displayed by analogous models where only the logistic form is considered.

\subsection{Predator-prey model}

Let us consider a two-species system with one prey, $y$, and one predator, $x$. For the prey, we propose a logistic demography and a predation term with saturation, as in \cite{murray}. Further, we assume that the dynamics of the predator presents the Allee effect, with a threshold of critical population, $x_t$, and a carrying capacity proportional to the prey population, $K_x=cy$. The equations read:   
\begin{subequations}
\begin{align}
\dot{y} &= r\,y \left(1-\frac{y}{K}\right) -A \frac{x y}{y+B}, \\
\dot{x} &= s \frac{x}{4cy} \left((cy-x_t)|cy-x_t|-(2x-x_t-cy)^2\right).
\end{align}
\label{eq:allee2}
\end{subequations}

When $x_t<y$ the equations become: 
\begin{subequations}
\begin{align}
\dot{y} &= r\,y \left(1-\frac{y}{K}\right) -A \frac{x y}{y+B},\\
\dot{x} &= s\,x (x-x_t) \left(1-\frac{x}{cy}\right), 
\end{align}
\end{subequations}
where the predator equation has the Nagumo form, 
and when $x_t>y$ we have:
\begin{subequations}
\begin{align}
\dot{y} &= r\,  y (1-y) -A \frac{x y}{y+B},  \\
\dot{x} &= -s\frac{x}{2cy} \left( (x-x_t)^2+(cy-x)^2 \right). \label{eq:alle1b}
\end{align}
\label{eq:alle1}
\end{subequations}

While a deconstruction of the polynomial expression of Eq.~(\ref{eq:alle1b}) in terms of biological facts can be hard, it is interesting to notice  that its functional form presents sensible features. Once the carrying capacity of the environment falls below the critical population size, the predator population should evolve towards extinction. In fact, this is what happens, as the derivative is always negative. But interestingly, there is still a ghost behavior remembering the existence of a critical size, as it is usual in saddle-node transitions. The speed towards extinction is not constant and depends on $x$. The rate of extinction is minimum when the population is at the critical size, and it increases for larger or smaller populations. 

\begin{figure}[t!]
\includegraphics[width=\columnwidth]{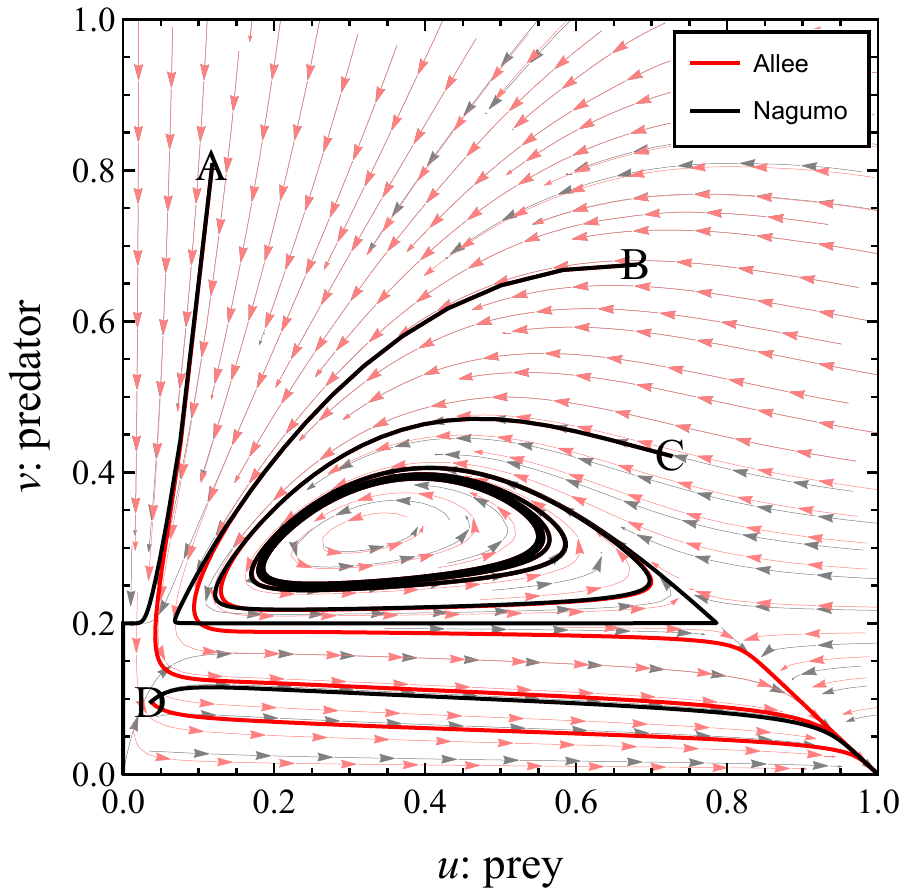}
\includegraphics[width=\columnwidth]{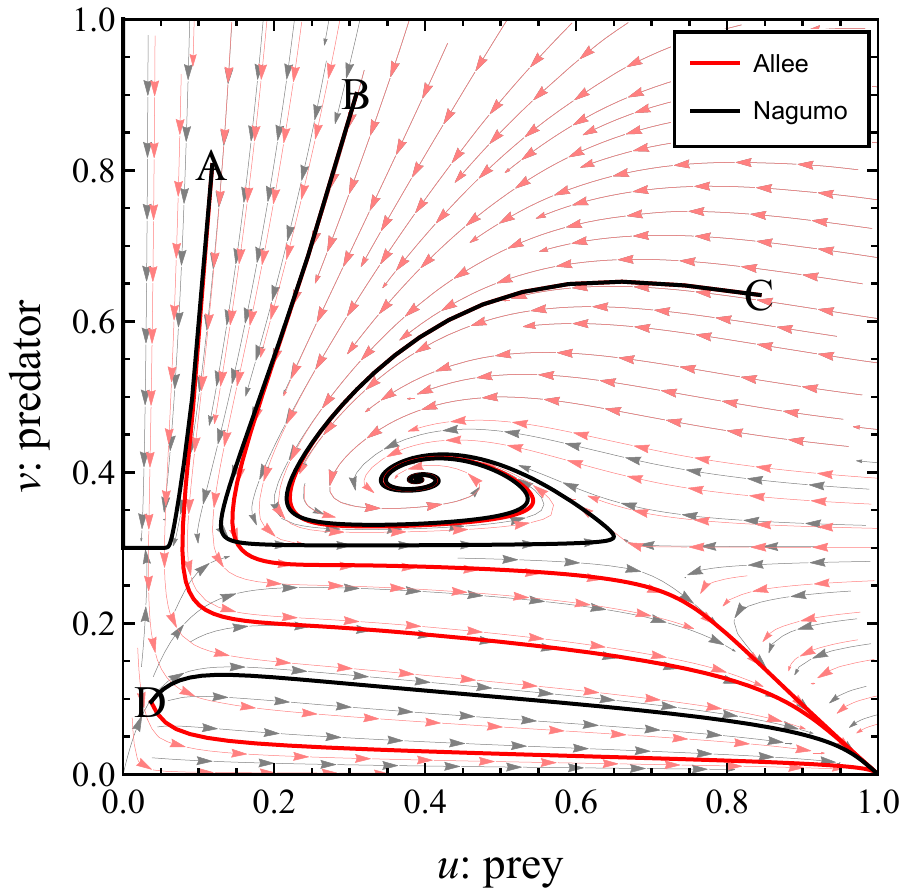}
\caption{Two examples of different behaviors displayed by the Nagumo (black) and the Allee (red) predator-prey models under the same conditions. Each panel shows the phase space, with a stream plot of the velocity field and four sets of initial conditions and the corresponding trajectories, as described in the text. Top: $b=0.15$, $v_t=0.2$; bottom: $b=0.25$, $v_t=0.3$; both: $a=\sigma=1$.}
\label{fig:predprey1}
\end{figure}

Let us rewrite  Eqs.~(\ref{eq:allee2}) in dimensionless form with the following change of variables and parameters:
\begin{gather*}
u(\tau)=y(t)/K,~~~~~v(\tau)=x(t)/cK,~~~~~\tau=rt, \\
a=cA/r,~~~~~~b=B/K, \\
v_t=x_t/cK,~~~~~~\sigma = scK/r.
\end{gather*}
Equations~(\ref{eq:allee2}) become:
\begin{subequations}
\begin{align}
\dot{u} &= u (1-u) -a \frac{uv}{u+b}, \\
\dot{v} &= \frac{\sigma v}{4u} \left( (u-v_t)|u-v_t|-(2v-v_t-u)^2\right).
\end{align}
\end{subequations}
When $u>v_t$, these are analogous to Murray's realistic predator-prey models studied in \cite{murray}:
\begin{subequations}
\begin{align}
\dot{u} &= u (1-u) -a \frac{uv}{u+b}, \\
\dot{v} &=  \sigma v (1-\frac{v}{u}),  
\end{align}
\label{eq:pred-prey}
\end{subequations}
where the existence of a limit cycles is verified when: 
\begin{align*}
\sigma &< 
\left(a-Q\right) \frac{1+a+b-Q}{2a}, \\
&\text{with } Q = \left((1-a-b)^2+4b\right)^{1/2} .
\end{align*}

We show in Fig.~\ref{fig:predprey1} phase portraits of typical scenarios where the two models (with Nagumo in black and Allee in red) have different behaviors. Each panel shows stream plots, to help visualize the flow, as well as four selected trajectories, to show the disparity of the basins of attraction. Observe, for example, the top panel, which corresponds to a set of parameters where a limit cycle exists for both models. The initial condition A  corresponds to a situation where the Nagumo model drives the prey to extinction, and the predator persists. This is clearly an unrealistic situation, which arises from the interchange of stability of the equilibria (as explained in the Introduction), and the proper Allee model solves it. The initial condition B is in the basin of attraction of the cycle for the Nagumo model, but for the Allee one, it also goes to the extinction of the predator. Condition C, instead, goes to the cycle in both models (observe that the trajectories are almost indistinguishable. The initial condition D, finally, has both models driving the predator to extinction, with different transients. The bottom panel shows the equivalent picture for another set of parameters, one where we see a stable spiral of coexistence instead of the limit cycle (related to it by a Hopf bifurcation). We also see the same general behavior: the two models differ in the nature and stability of equilibria, as well as in their basins of attraction.

\subsection{Competitive interaction }
We will focus now in a competitive interaction between two species. The usual equations for the dynamics of the population of  two competing species $x_1$ and $x_2$ are \cite{murray}:
\begin{subequations}
\begin{align}
\dot{x_1} &= r_1 x_1 \left(1-\frac{x_1}{K_1}-a_{12}\frac{x_2}{K_1}\right),\\
\dot{x_2} &= r_2 x_2 \left(1-\frac{x_2}{K_2}-a_{21}\frac{x_1}{K_2}\right),
\end{align}
\label{comp1}%
\end{subequations}
that with a proper change of variables can be reduced to 
\begin{subequations}
\begin{align}
\dot{u_1} &=  u_1 (1-u_1-\alpha_{12} u_2),\\
\dot{u_2} &= \rho u_2 (1-u_2-\alpha_{21}u_1). 
\end{align}
\label{comp2}
\end{subequations}

These equations have four equilibria, whose stability will be discussed in the Appendix:
\begin{center}
    \begin{tabular}{ll}
    $(1,0)$,     & \hspace{1cm}$(0,1)$,  \\
    &\\
    $(0,0)$,     & \hspace{1cm} $\left(\displaystyle{\frac{1-\alpha_{12}}{1-\alpha_{12}\alpha_{21}},\frac{1-\alpha_{21}}{1-\alpha_{12}\alpha_{21}}}\right)$.
    \end{tabular}
\end{center}    

Equations~(\ref{comp2})  can be adapted to include the Allee effect but this adaptation should be done with care. 
Direct inclusion of a multiplying monomial to account for the Allee effect, as in Eq.~(\ref{eq:comp1}), and a proper non-dimensionalization of the system will leave us with: 
\begin{subequations}
\begin{align}
\dot{u_1} &=  u_1 (1-u_1-\alpha_{12} u_2) (u_1-t_1),\\
\dot{u_2} &= \rho u_2 (1-u_2-\alpha_{21}u_1)(u_2-t_2),  
\end{align}
\label{comp3}%
\end{subequations}
where now $t_i$ is the survival threshold for species $u_i$. However, these equations do not contemplate the situation when the depletion of the carrying capacity of the environment due to the presence of the competing species can lead one of them below its survival threshold. The consequences of this fact will be discussed later.

A correct way to include the Allee effect is, again, by reformulating the equations as:
\begin{subequations}
\begin{align}
\dot{u_1} &= u_1 \left[(1-\alpha_{12} u_2-t_1) |1-\alpha_{12} u_2-t_1|/4 \right.\nonumber\\ 
&~\left.-\left((u_1-t_1)-(1-\alpha_{12} u_2-t_1)/2)^2\right)\right],\\
\dot{u_2} &= u_2 \left[(1-\alpha_{21} u_1-t_2) |1-\alpha_{21} u_1-t_2|/4 \right.\nonumber \\
&~\left.-\left((u_2-t_2)-(1-\alpha_{21} u_1-t_2)/2)^2\right)\right]. 
\end{align}
\label{comp4}
\end{subequations}

When $(1-\alpha_{12} u_2-t_1)>0$, each equation adopts the form: 
\begin{equation}
\dot{u_i}=  u_i (1-u_i-\alpha_{ij} u_j) (u_i-t_i),
\end{equation}
and in the other case:
\begin{equation}
\dot{u_i}=-\frac{u_i}{2} \left(
   (1-\alpha_{ij} u_j-u_i)^2+(t_i-u_i)^2\right).
\end{equation}
As in the predator-prey model, the derivative in this last case is always negative, leading to the extinction of the species, as it should.

In both cases, Eqs.~(\ref{comp3}) and (\ref{comp4}), the consideration of the Allee effect affecting two competitive species adds five new equilibria to the already present in Eqs.~(\ref{comp2}). These five equilibria  are: 
\begin{center}
    \begin{tabular}{ll}
    $(t_1,0)$,     & \hspace{1cm}$(0,t_2)$,  \\
    &\\
    $(t_1, 1- \alpha_{21} t_1)$,     & \hspace{1cm} $(1-\alpha_{12}t_2,t_2)$,\\
    &\\
    $(t_1,t_2)$.&
    \end{tabular}
\end{center}    

\begin{figure}[t!]
\includegraphics[width=\columnwidth]{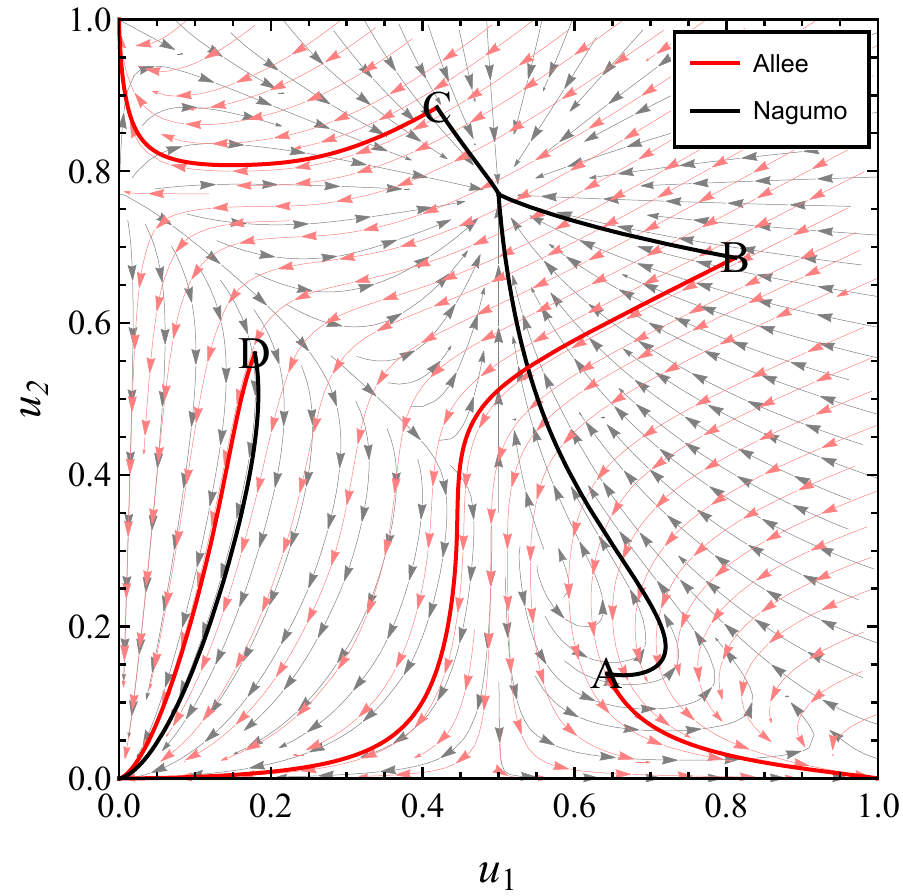}
\includegraphics[width=\columnwidth]{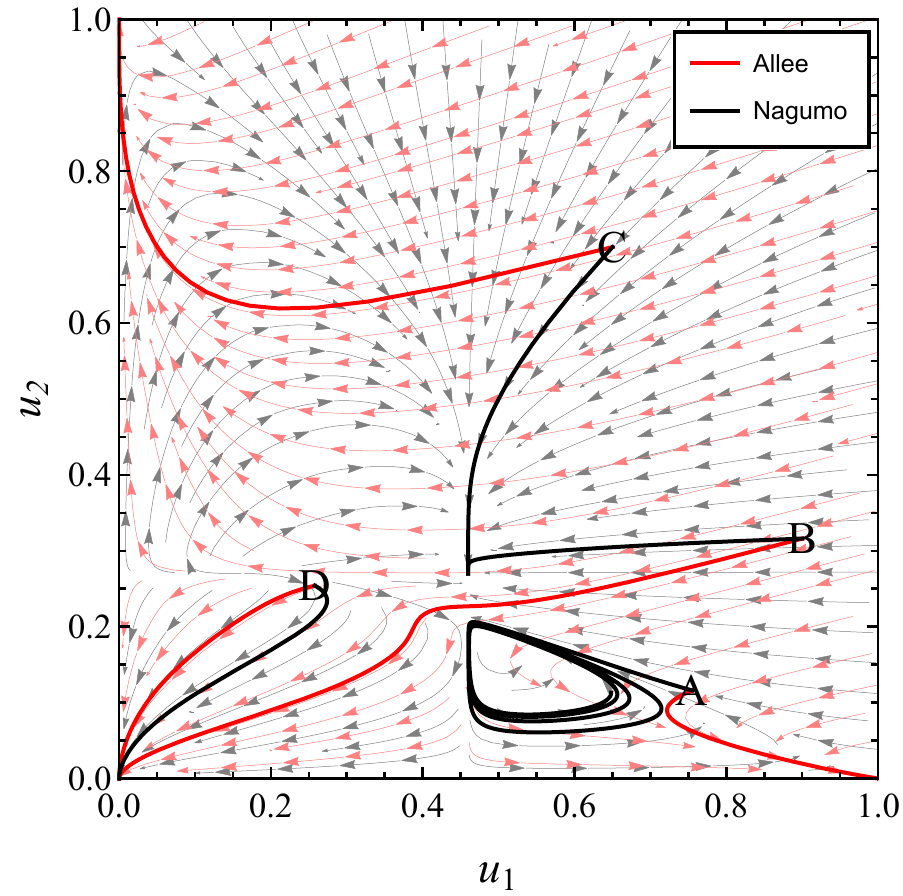}
\caption{Two examples of different behaviors displayed by the Nagumo (black) and the Allee (red) competition models under the same conditions. Each panel shows the phase space, with a stream plot of the velocity field and four sets of initial conditions and the corresponding trajectories, as described in the text. 
Top: $\alpha_{12}=1.61$, $\alpha_{21}=0.13$, $t_1=0.5$, $t_2=0.77$; bottom: $\alpha_{12}=3.1$, $\alpha_{21}=1.3$, $t_1=0.46$, $t_2=0.27$.}
\label{fig:comp1}
\end{figure}

We start by discussing the differences between Eqs.~(\ref{comp2}) and Eqs.~(\ref{comp3}) and (\ref{comp4}) regarding the common equilibria.
As shown in \cite{murray}, (0,0) is an unstable equilibrium of Eqs.~(\ref{comp2}). Due to the Allee effect, extinction is now allowed and thus (0,0) becomes stable. This can be clearly appreciated in the trajectories B and D of Figs.~\ref{fig:comp1}. Meanwhile, in all the cases, the stability of (1,0) and (0,1) is only defined by the values of $\alpha_{ij}$. This is apparent when we calculate the Jacobian evaluated at those equilibria (see the Appendix).
A second difference emerges around the fourth equilibrium.
While a sufficient condition for it to exist when considering Eqs.~(\ref{comp2}) and (\ref{comp3}) is that $\alpha_{ij}<1$,  when considering Eqs.~(\ref{comp4}) we also need that the equilibrium values are both greater than the  corresponding thresholds, $t_1$ and $t_2$. The former, being not a limitation for Eqs.~(\ref{comp3}), leads to an unrealistic oscillatory solution around the equilibrium values as shown in  the trajectory A of Fig.~\ref{fig:comp1} (bottom). 

It is in fact the disappearance of some equilibria for certain parameter values what establishes the difference between Eqs.~(\ref{comp3}) and (\ref{comp4}). The cubic terms introduce new nullclines, responsible for the new five equilibria,  which behave differently in Eqs.~(\ref{comp3}) and (\ref{comp4}).  If the dynamics of a  population drives its number below the survival threshold, the corresponding population should get extinct. At this moment there should be a  saddle-node bifurcation that eliminates all the equilibria with the exception of those where the extinction of the corresponding solution is predicted. However, Eq.~(\ref{comp3}) erroneously predicts a transcritical bifurcation in which the threshold value turns into a stable equilibrium. This fact induces non-realistic behaviors. These include the above-mentioned oscillations and the stabilization of some of the equilibria in which the values $t_1$ and $t_2$ are involved. If we observe Figs. \ref{fig:comp1} we can note the differences between trajectories starting from initial conditions A, B, and C that in the case of the Nagumo model end in a non-realistic equilibrium or behavior (limit cycle). 
As expected, these stable equilibria are not present when a proper formulation, Eqs.~(\ref{comp4}), is considered.  The additional nullclines trace new limits for the basin of attraction of the old (and new) equilibria, resulting in a temporal behavior dependent on the initial conditions. The last was only the case for $\alpha_{ij}>1$ when considering Eqs.~(\ref{comp2}).

\begin{figure}
\includegraphics[width=\columnwidth]{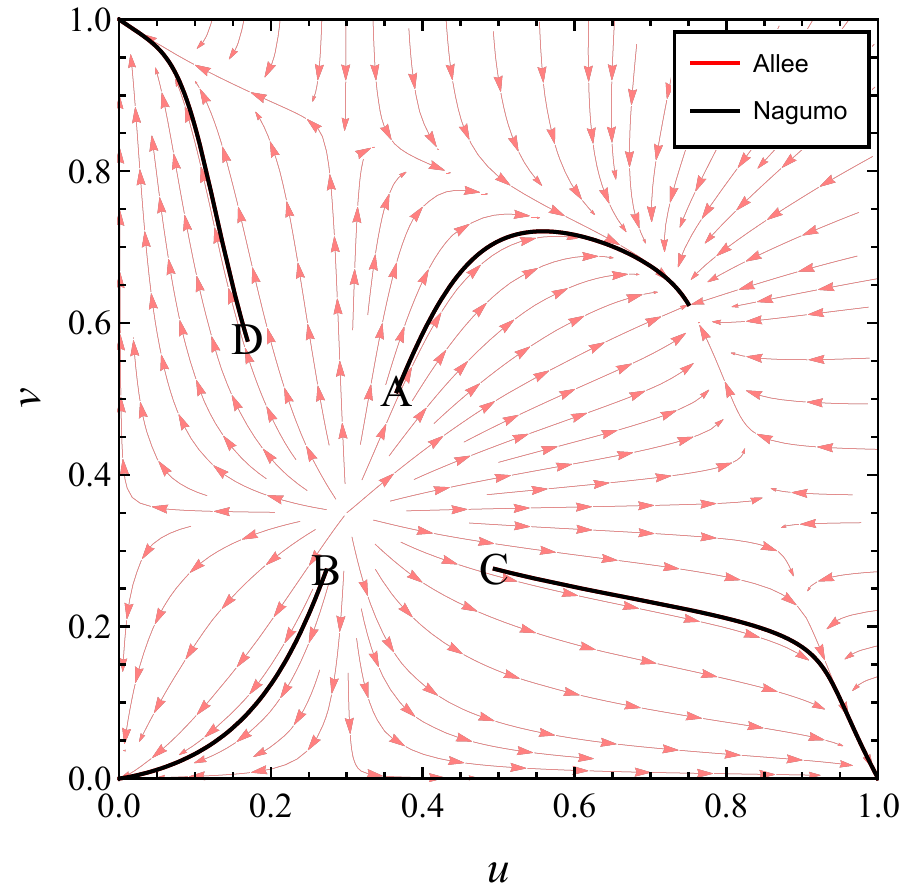}
\caption{Phase space and trajectories for different initial conditions, when coexistence is possible.  Both the flow and the trajectories coincide for both models, hence only one color is visible in the plots. Parameters: $\alpha_{12}=0.4$, $\alpha_{21}=0.5$, $t_1=0.3$, $t_2=0.35$.}
\label{fig:comp-coin}
\end{figure}

It is also important to note that, within the range of validity of Eqs.~
(\ref{comp3}),  the dynamic behavior predicted by both equations is exactly the same. This is the case when the coexistence solution is stable. We show in Fig. \ref{fig:comp-coin}  exemplary trajectories starting at  different conditions. In all the cases the trajectories are overlapped. In particular, initial condition A ends in the coexistence equilibrium. 


\section{Conclusion}
The inclusion of the Allee effect when studying population dynamics is more than a subtle detail. In many circumstances, like the ones described in the present article, the Allee effect can be responsible for the extinction of a population, otherwise supposed to survive no matter how low its population falls.
A mathematical description of the Allee effect, especially when dealing with interacting populations, should be done with care. The main reason is that, while the threshold for survival depends mostly on the population under study, the carrying capacity of the environment can suffer from variations that can lead to a degrading and non-sustainable situation.
Once a population number is driven by environmental factors below its critical threshold, the collapse is unavoidable.
In this work, we have presented two examples of interacting populations: a predator-prey system and a competitive situation. In both cases, we have considered the Allee effect and constructed the respective equations to mathematically describe the dynamic behavior of the populations subject to this effect. We have shown that Nagumo-kind models, with a third-degree polynomial, which are usual in single population modeling, give meaningless results for interacting populations. We also showed a correct way to model the strong Allee effect, in the sense that the ecologically sensible dynamics can be studied without confusion.
The study of the correct equations allowed us to understand the true consequences of including the Allee effect, reflected in the stabilization of the extinction of the species and in the strong dependence of the steady-state on the initial conditions. The latter occurs due to the appearance of new basins of attraction with respect to the models lacking the Allee effect. 
In this work we studied systems of two interacting species. The presence of multiple species, with different interacting relationships among them, induce more complex behaviors even in the absence of the Allee effect \cite{kuposc}. Considering its inclusion in those systems is certainly worth studying, and will be the subject of future works.

\section*{Acknowledgments}
The authors acknowledge support from CONICET [PIP 112-2017-0100008 CO], UNCUYO [SIIP 06/C546] and ANPCyT [PICT-2018-01181].

\newpage
\begin{strip}
\appendix
\section{Linear stability analysis}

Here we analyze the stability of the equilibria corresponding to the predator-prey and competition systems, by studying the eigenvalues of the corresponding Jacobian matrices evaluated at each equilibrium.

\subsection{Predator-prey model}
The Jacobian of Eqs.~(\ref{eq:pred-prey}),  which is  valid when $u>v_t$ is:
\begin{equation*}
J(u,v)= 
\begin{pmatrix}
	1 - 2 u - \frac{a b v}{b + u}^2 & -\frac{a u}{b + u}\\
	& \\
	\frac{\sigma v^2 (v - v_t)}{u^2} &-\frac{\sigma}{u} (3 v^2-2 u v+ (u-2v) v_t)
	\end{pmatrix}.
\end{equation*} 

We evaluate the Jacobian at each equilibrium.

\bigskip	
1)	
\begin{equation*}
\lim_{(u,v) \to (0,0)} J(u,v)=\begin{pmatrix}
1&0\\
& \\
0&0 
\end{pmatrix}.
\end{equation*} 
Both eigenvalues are real and non negative, so (0,0) is  unstable.

\medskip
2)
\begin{equation*}
J(1,0)=\begin{pmatrix}
-1&-\frac{a}{1+b}\\
& \\
0&-\sigma v_t 
\end{pmatrix}.
\end{equation*}
Since $v_t>0$,  $(1,0)$ is a stable node.

\medskip
3)
This case corresponds to $u_1=v_1=\frac{1}{2} \left(\sqrt{(a+b-1)^2+4 b}-a-b+1\right)$.
\begin{equation*}
J(u_1,v_1)=\begin{pmatrix}
1-u_1\left(2+\frac{ab}{(b+u_1)^2}\right)   &-\frac{au_1}{b+u_1}\\
& \\
\sigma (u_1-v_t)&-\sigma (u_1-v_t)
\end{pmatrix}.
\end{equation*}

This case presents multiple behaviors. Depending on the combination of parameter values the equilibrium can be a stable node, a stable spiral, a saddle node or an unstable spiral enclosed by a limit cycle. 

\medskip
4)
This case corresponds to $u_2=\frac{1}{2} \left(-\sqrt{(b+1)^2-4 a v_t}-b+1\right)$.
\begin{equation*}
J(u_2,u_2)=\begin{pmatrix}
1-2u_2-\frac{abv_t}{(b+u_2)^2}&-\frac{au_2}{b+u_2}\\
& \\
0&\sigma v_t \left(1-\frac{v_t}{u_2} \right)
\end{pmatrix}.
\end{equation*}

\end{strip}

\begin{strip}
A careful numerical analysis of the eigenvalues of this Jacobian shows that the equilibrium is always a saddle.

\medskip
5)
This case corresponds to $u_3=\frac{1}{2} \left(\sqrt{(b+1)^2-4 a v_t}-b+1\right)$.
\begin{equation*}
J(u_3,u_3)=\begin{pmatrix}
1-2u_3-\frac{abv_t}{(b+u_3)^2}&-\frac{au_3}{b+u_3}\\
& \\
0&\sigma v_t \left(1-\frac{v_t}{u_3} \right)
\end{pmatrix}.
\end{equation*}

This equilibrium can be a stable node or a saddle.

\bigskip
When $u<v_t$ the only equilibra are (1,0) (stable) and (0,0) (unstable).  In this case the Jacobian is 
\begin{equation*}
J(u,v)=\begin{pmatrix}
	1 - 2 u - \frac{a b v}{b + u}^2 & -\frac{a u}{b + u}\\
	& \\
	 \frac{\sigma v}{2 u^2}
   \left(-u^2+v^2+(v-v_t)^2\right)&-\frac{s}{2 x} \left((u-v)^2-u v+(2v-v_t)^2\right)
	\end{pmatrix}.
	\end{equation*} 
 
1)	
\begin{equation*}
\lim_{(u,v) \to (0,0)} J(u,v)=\begin{pmatrix}
1&0\\
& \\
0&0 
\end{pmatrix}.
\end{equation*} 
Both eigenvalues are real and non negative, so (0,0) is  unstable.
\vspace{.5cm}

2)
\begin{equation*}
J(1,0)=\begin{pmatrix}
-1&-\frac{a}{1+b}\\
& \\
0&-\frac{\sigma}{2} (1+v_t^2)
\end{pmatrix}.
\end{equation*}
Hence $(1,0)$ is then a stable node.

\subsection{Competition model}
The Jacobian of Eqs.~(\ref{comp3}), which is also valid for Eqs.~(\ref{comp4}) when $(1-\alpha_{ij} u_j)>t_i$, is: 
\begin{equation*}
J(u_1,u_2)=\begin{pmatrix}
	u_1(2-3u_1-2 \alpha_{12} u_2) + t_1(2x_1+ \alpha_{12} u_2-1) & \alpha_{12} (t_1 - u_1) u_1\\
	& \\
	\alpha_{21} x_2 (t_2 - u_2) & u_2(2-3u_2-2 \alpha_{21} u_1) + t_2(2x_2+ \alpha_{21} u_1-1) 
	\end{pmatrix}.
	\end{equation*} 
	
We evaluate the Jacobian at each equilibrium.

\bigskip
1)	
\begin{equation*}
J(0,0)=\begin{pmatrix}
-t_1&0\\
& \\
0&-t_2 
\end{pmatrix}.
\end{equation*} 
Both eigenvalues are real and negative, so (0,0) is a stable node.

\medskip
2)
\begin{equation*}
J(1,0)=\begin{pmatrix}
t_1-1&(t_1-1)\alpha_{12}\\
& \\
0&t_2(\alpha_{21}-1) 
\end{pmatrix}.
\end{equation*} 
Since $t_1<1$,  $(1,0)$ is a stable (unstable) node if  $\alpha_{21}<1$ ($\alpha_{21}>1$).

\end{strip}

\newpage
\noindent 3)
\begin{strip}
\begin{equation*}
J(0,1)=\begin{pmatrix}
0&t_1(\alpha_{12}-1)\\
& \\
(t_2-1)\alpha_{21}&t_2-1 
\end{pmatrix}.
\end{equation*} 
This case is analogous to the previous one.

\medskip
4)

Here we analize the case  $u_1=\displaystyle{\frac{1-\alpha_{12}}{1-\alpha_{12}\alpha_{21}}}$ and  $u_2=\displaystyle{\frac{1-\alpha_{21}}{1-\alpha_{12}\alpha_{21}}}$.
\begin{equation*}
J(u_1,u_2)=\frac{1}{(1- \alpha_{12} \alpha_{21})^2} 
\begin{pmatrix}
( \alpha_{12}-1) (1 -t_1 +\alpha_{12} (\alpha_{21} t_1-1))&(\alpha_{12}-1) \alpha_{12} (1 - t1 + \alpha_{12}(\alpha_{21} t_1-1))\\
& \\
(\alpha_{21}-1) \alpha_{21} (1 - t_2 + \alpha_{21} (\alpha_{12} t_2-1))&( \alpha_{21}-1) (1 - t_2 + \alpha_{21} (\alpha_{12} t_2-1))
\end{pmatrix}.
\end{equation*} 
The analysis of this case is not straightforward as there are four parameters but together with an exploration of the phase space we can observe that the equilibrium is stable provided that $\alpha_{12}<1$ and $\alpha_{21}<1$. Also, it is necessary to fulfill the conditions $u_1>t_1$ and $u_2>t_2$.

\medskip
5)
\begin{equation*}
J(t_1,t_2)=\begin{pmatrix}
t_1 (1 - t_1 - \alpha_{12} t_2)&0\\
& \\
0&t_2 (1 - t_2 - \alpha_{21} t_1)
\end{pmatrix}.
\end{equation*} 
Both eigenvalues are real. In order to have a stable node at $(t_1,t_2)$, we need  $(1-\alpha_{ij} u_j)<t_i$, otherwise it is unstable. We recall that when $(t_1,t_2)$ is an equilibrium of Eqs.~(\ref{comp4}), it is unstable since
the condition $(1-\alpha_{ij} u_j)>t_i$ holds.

\medskip
6)
\begin{equation*}
J(t_1,0)=\begin{pmatrix}
t_1 (1 - t_1)&0\\
& \\
0&(1 - \alpha_{21} t_1) 
\end{pmatrix}.
\end{equation*} 
Since $t_1<1$, $(t_1,0)$ is not stable.

\medskip
7)
\begin{equation*}
J(0,t_2)=\begin{pmatrix}
(1 - \alpha_{12} t_2) &0\\
& \\
0&t_2 (1 - t_2)
\end{pmatrix}.
\end{equation*} 
This case is analogous to the previous one.

\medskip
8)
\begin{equation*}
J(t_1,1-\alpha_{21} t_1)=\begin{pmatrix}
t_1 (1 - t_1 - \alpha_{12}(1-\alpha_{21} t_1))
&0\\
& \\
- \alpha_{21}(1 - \alpha_{21} t_1) (1 - \alpha_{21} t_1 - t_2)&(1 - \alpha_{21} t_1) (1 - \alpha_{21} t_1 - t_2)
\end{pmatrix}.
\end{equation*} 
While both eigenvalues are always real, they can be positive or negative depending on the values adopted by the parameters. 

\medskip
9)
\begin{equation*}
J(1-\alpha_{12} t_2,t_2)=\begin{pmatrix}

(1 - \alpha_{12} t_2) (1 - \alpha_{12} t_2 - t_1)&-\alpha_{12}(1 - \alpha_{12} t_2) (1 - \alpha_{12} t_2 - t_1)\\
& \\
0&t_2 (1 - t_2 - \alpha_{21}(1-\alpha_{12} t_2))
\end{pmatrix}.
\end{equation*} 
This case is analogous to the previous one.


If we  consider Eqs.~(\ref{comp4}) when $(1-\alpha_{ij} u_j)<t_i$, the only equilibrium  is $(0,0)$ and the Jacobian is: 
\begin{equation*}
K(0,0)=\frac{1}{2}\begin{pmatrix}
-(1+ t_1^2) &0\\
& \\
0&-(1+ t_2^2)
\end{pmatrix}.
\end{equation*}
Both eigenvalues are real and negative, so $(0,0)$ is a stable node.

\medskip
The last case is when the carrying capacity of one of the species is above its threshold (for example $u_1$) and the other below it.
In this case $(1-\alpha_{12} u_2)>t_1$ and $(1-\alpha_{21} u_1)\leq t_2$.
There are three equilibria: $(1,0)$, $(0,0)$ and ($t_1$,0), though $(0,0)$ is just a limiting case because in that case, as $u_1=0$ then $t_2=1$.
The Jacobian matrix evaluated at each equilibrium follows.

\medskip
1)
\begin{equation*}
 L(0,0)=\begin{pmatrix}
	- t_1&0\\
	& \\
	0&-\frac{1}{2}(1+ t_2^2)
\end{pmatrix}.
\end{equation*}
Both eigenvalues are real and negative, so $(0,0)$ is a stable node.

\medskip
2)
\begin{equation*}
L(1,0)=\begin{pmatrix}
 t_1-1&\alpha_{12}(t_1-1)\\
& \\
0&-\frac{1}{2}\left((\alpha_{21}-1)^2+t_2^2\right)
\end{pmatrix}.
\end{equation*}
Since $t_1<1$, both eigenvalues are real and negative, hence $(1,0)$ is a stable node.

\medskip
3)
\begin{equation*}
L(t_1,0)=\begin{pmatrix}
t_1(1-t_1)&0\\
& \\
0&-\frac{1}{2}\left((\alpha_{21}-1)^2+t_2^2\right)
\end{pmatrix}.
\end{equation*}
Since $t_1<1$, $(t_1,0)$ is a saddle.

\bigskip
This completes the linear stability analysis.
\end{strip}

\end{document}